\documentclass{PoS}
\usepackage{psfig,epsfig,colordvi}

\title{2-loop additive mass renormalization with clover fermions and
  Symanzik improved gluons}

\ShortTitle{Additive Mass Renormalization}

\author{\speaker{Apostolos Skouroupathis} \\
        Department of Physics, University of Cyprus, 
        Nicosia, CY-1678, Cyprus\\
        E-mail: \email{php4as01@ucy.ac.cy}}

\author{Martha Constantinou\\
        Department of Physics, University of Cyprus, 
        Nicosia, CY-1678, Cyprus\\
        E-mail: \email{phpgmc1@ucy.ac.cy}}

\author{Haralambos Panagopoulos\\
        Department of Physics, University of Cyprus, 
        Nicosia, CY-1678, Cyprus\\
        E-mail: \email{haris@ucy.ac.cy}}

\abstract{We calculate the critical value of the hopping parameter,
$\kappa_c$, in Lattice QCD, up to two loops in perturbation theory.
We employ the Sheikholeslami-Wohlert (clover) improved action for
Wilson fermions and the Symanzik improved gluon action for 4- and
6-link loops.

The quantity which we study is a typical case of a vacuum
expectation value resulting in an additive renormalization; as such,
it is characterized by a power (linear) divergence in the lattice
spacing, and its calculation lies at the limits of applicability of
perturbation theory.

Our results are polynomial in $c_{\rm SW}$ (clover parameter) and
cover a wide range of values for the Symanzik coefficients $c_i$.
Furthermore, the dependence on the number of colors $N$ and the
number of fermionic flavors $N_f$ is shown explicitly. In order to
compare our results to non perturbative evaluations of $\kappa_c$
coming from Monte Carlo simulations, we employ an improved
perturbation theory method applied to improved actions.}

\FullConference{XXIV International Symposium on Lattice Field
Theory\\
         July 23-28 2006\\
         Tucson Arizona, US}

\begin{document}
\def\slashed{{/}\mskip-10.0mu}
\def\pcircslash{\slashed {p\mskip -5mu ^{^\circ}}}

\section{Introduction}
In the present work, we study clover fermions~\cite{SW} and Symanzik improved
gluons~\cite{Sym}, for the calculation of the critical value of the hopping
parameter, $\kappa_c$ in perturbative Lattice QCD, to two loops
order in a perturbative expansion.

Concerning gauge fields, we employ the Symanzik improved action
~\cite{Sym}. The purpose of introducing such an improvement of the
lattice action, is to minimize the consequences of a non-zero
lattice spacing, order by order in perturbation theory. Several
choices are made for the coefficients of the action corresponding to
values which are most often used in the literature.

The discretization of the theory via lattice regularization
introduces some difficulties that do not exist in the continuum
theory. To be able to recover the continuum limit, we must demand
strict locality and absence of doublers. It is known that these
requirements lead to breaking of chiral symmetry. Of course, while
approaching the continuum limit we expect to recover chirality. This
is the point where one must introduce the hopping parameter. The
main idea is to ensure chiral symmetry, by setting the renormalized
fermionic mass ($m_R$) equal to zero. Because of additive
renormalization, setting the bare fermionic mass equal to zero is
not enough. Hence, there is a critical mass, the role of which is to
guarantee that $m_R$ vanishes. This quantity is directly related to
the hopping parameter $\kappa$. Its critical value ($\kappa_c$) is
responsible for restoring chiral symmetry.

Previous works on the hopping parameter and its critical value
appear in the literature for Wilson fermions-plaquette action
gluons~\cite{FP} and for clover fermions-plaquette action
gluons~\cite{PP}. The procedure and notation in our work is the
same as in the above references.

Our results for $\kappa_c$ (and thus for the critical
mass) depend on the number of colors ($N$) and fermionic flavors
($N_f$). Besides that, there is an 
explicit dependence on the clover parameter $c_{\rm SW}$ which 
is kept as a free parameter. The
dependence of the results on the choice of Symanzik coefficients
cannot be given in closed form; instead, we present it in a list of
Tables and Figures.

The rest of the paper is organized as follows: In Sec.~\ref{sec2} we
define the actions in their discretized form, as well as the
connection between the hopping parameter and fermionic mass.
Furthermore, there is a description of our calculations along with
the necessary Feynman diagrams. In Sec.~\ref{sec3} our results are
presented and compared with previous Monte Carlo simulations.
Finally, in Sec.~\ref{sec4} we apply to our 1- and 2-loop results an
improvement method, proposed by us \cite{cactus1,cactus2,CPS}. This
method resums a certain infinite class of subdiagrams, to all orders
in perturbation theory, leading to improved results. A full write-up
of this work, including detailed tables of results is forthcoming~\cite{SCP}.

\section{Formulation of the Problem}
\label{sec2}

We begin with the Wilson formulation of the QCD action on the
lattice, with the addition of the clover (SW)~\cite{SW} term for
fermions. In standard notation, it reads:
\begin{eqnarray}
S_L &=& S_G + \sum_{f}\sum_{x} (4r+m)\bar{\psi}_{f}(x)\psi_f(x)
\nonumber \\
&-& {1\over 2}\sum_{f}\sum_{x,\,\mu} \bigg{[}\bar{\psi}_{f}(x)
\left( r - \gamma_\mu\right)
U_{\mu}(x)\psi_f(x+{\mu}) \nonumber + \bar{\psi}_f(x+{\mu})\left( r + \gamma_\mu\right)U_{\mu}(x)^\dagger\psi_{f}(x)\bigg{]}\nonumber \\
&+& {i\over 4}\,c_{\rm SW}\,\sum_{f}\sum_{x,\,\mu,\,\nu}
\bar{\psi}_{f}(x) \sigma_{\mu\nu} {\hat F}_{\mu\nu}(x) \psi_f(x),
\label{latact}
\end{eqnarray}
\begin{eqnarray}
\hskip -8mm \hbox {where:}\,\, {\hat F}_{\mu\nu} &\equiv& {1\over{8a^2}}\,
(Q_{\mu\nu} - Q_{\nu\mu})\\
\hskip -8mm \hbox {and:}\,\, Q_{\mu\nu} &=& U_{x,\, x+\mu}U_{x+\mu,\, x+\mu+\nu}U_{x+\mu+\nu,\, x+\nu}U_{x+\nu,\, x}+ U_{ x,\, x+ \nu}U_{ x+ \nu,\, x+ \nu- \mu}U_{ x+ \nu- \mu,\, x- \mu}U_{ x- \mu,\, x} \nonumber \\
&+& U_{ x,\, x- \mu}U_{ x- \mu,\, x- \mu- \nu}U_{ x- \mu- \nu,\, x- \nu}U_{ x- \nu,\, x}+ U_{ x,\, x- \nu}U_{ x- \nu,\, x- \nu+ \mu}U_{ x- \nu+ \mu,\, x+
\mu}U_{ x+ \mu,\, x}
\label{latact2}
\end{eqnarray}

The clover coefficient $c_{\rm SW}$ is treated here as a free
parameter; $r$ is the Wilson parameter; $f$ is a flavor index;
$\sigma_{\mu\nu} =(i/2) [\gamma_\mu,\,\gamma_\nu]$. Powers of the
lattice spacing $a$ have been omitted.

Regarding gluons, we use the Symanzik action,
involving Wilson loops with 4 and 6 links:
\begin{eqnarray}
S_G&=&\frac{2}{g^2} \Bigg[ c_0 \sum_{\rm plaq} {\rm Re\,
 Tr\,}(1-U_{\rm plaq}) + c_1 \sum_{\rm rect} {\rm Re \, Tr\,}(1- U_{\rm rect}) \nonumber \\
&&\quad\, +c_2 \sum_{\rm chair} {\rm Re\, Tr\,}(1- U_{\rm
chair})+c_3 \sum_{\rm paral} {\rm Re \,Tr\,}(1- U_{\rm paral})\Bigg]
\label{gluonaction}
\end{eqnarray}
The correct classical continuum limit requires: $c_0 + 8 c_1 + 16 c_2 + 8
c_3 = 1$.
The full action is: $S = S_L + S_{gf} + S_{gh} + S_m.$, where
$S_{gf}$, $S_{gh}$, $S_m$ are standard gauge fixing, ghost and
measure terms.

The bare fermionic mass $m_B$ must be set to zero for chiral
invariance in the classical continuum limit. The value of the
parameter $c_{\rm SW}$ and of the Symanzik coefficients $c_i $ can
be chosen arbitrarily; they are normally tuned in a way as to
minimize ${\cal O}(a)$ effects. Terms proportional to $r$ in the
action, as well as the clover terms, break chiral invariance. They
vanish in the classical continuum limit; at the quantum level, they
induce nonvanishing, flavor-independent corrections to the fermion
masses. Numerical simulation algorithms usually employ the hopping
parameter,
\begin{equation}
\kappa\equiv{1\over 2\,m_B\,a + 8\,r}
\end{equation}
as an adjustable quantity. Its critical value, at which chiral
symmetry is restored, is thus $1/8r$ classically, but gets shifted
by quantum effects.

We denote by $dm$ the perturbative contribution that must be added
to the bare mass, in order to lead to zero renormalized mass. At
tree level, $m_B=0$.

\begin{equation}
dm=dm_{\rm (1-loop)}+dm_{\rm (2-loop)}
\label{Totaldm}
\end{equation}

Two diagrams contribute to $dm_{\rm (1-loop)}$, shown in Figure 1.
The quantity $dm_{\rm (2-loop)}$ receives contributions from a total
of 26 diagrams, shown in Figure 2. Genuine 2-loop diagrams must 
be evaluated at $m_B\to 0$; in addition, one must include to this
order the 1-loop diagram containing an ${\cal O}(g^2)$ mass
counterterm (diagram 23). Certain sets of diagrams, corresponding to
renormalization of 1-loop propagators, must be evaluated together in
order to obtain an infrared-convergent result: These are diagrams
7+8+9+10+11, 12+13, 14+15+16+17+18, 19+20, 21+22+23.

\section{Numerical Results}
\label{sec3}

We have selected a set of most widely used values for the Symanzik
coefficients, shown in Table 1.
\begin{table}
\begin{center}
\begin{tabular}{lr@{}lr@{}lr@{}l}
\hline \hline
\multicolumn{1}{c}{Action}&
\multicolumn{2}{c}{$c_0$}&
\multicolumn{2}{c}{$c_1$} &
\multicolumn{2}{c}{$c_3$} \\
\hline \hline
Plaquette               &  1&.0             &  0&.0              &  0&.0        \\
Symanzik                &  1&.6666667       & -0&.083333         &  0&.0        \\
TILW, $\beta c_0=8.60$  &  2&.3168064       & -0&.151791         & -0&.0128098  \\
TILW, $\beta c_0=8.45$  &  2&.3460240       & -0&.154846         & -0&.0134070  \\
TILW, $\beta c_0=8.30$  &  2&.3869776       & -0&.159128         & -0&.0142442  \\
TILW, $\beta c_0=8.20$  &  2&.4127840       & -0&.161827         & -0&.0147710  \\
TILW, $\beta c_0=8.10$  &  2&.4465400       & -0&.165353         & -0&.0154645  \\
TILW, $\beta c_0=8.00$  &  2&.4891712       & -0&.169805         & -0&.0163414  \\
Iwasaki                 &  3&.648           & -0&.331            &  0&.0        \\
DBW2                    & 12&.2688          & -1&.4086           &  0&.0        \\
\hline \hline
\end{tabular}
\caption{Input parameters $c_0$, $c_1$, $c_3$.} \label{tab1}
\end{center}
\end{table}
(In all these cases, $c_2=0$. In general, for given values of
$C_1\equiv c_2+c_3$, $C_2\equiv c_1-c_2-c_3$ the dependence on $c_2$
is polynomial and thus we need not choose a numerical value for it.)

\noindent
\begin{minipage}{0.3\linewidth}
\begin{center}
{\centerline{\epsfig{figure=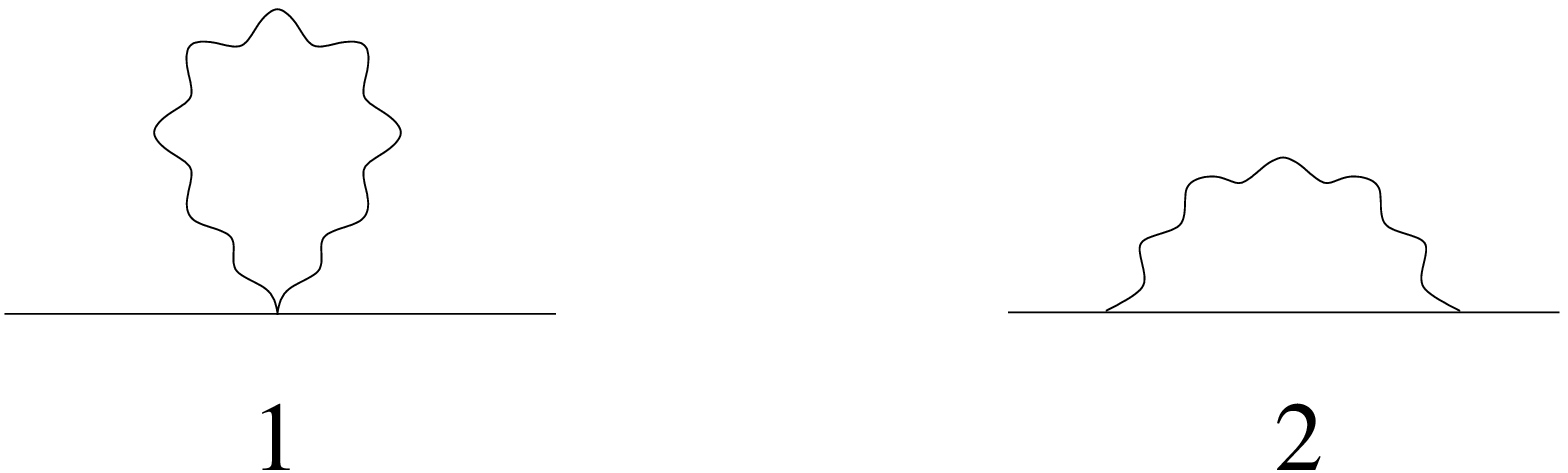,height=1.2truecm}}}
\end{center}
{\small {\bf Figure 1:} One-loop diagrams contributing to $dm_{\rm (1-loop)}$.
Wavy (solid) lines represent gluons (fermions).}
\end{minipage}\hskip0.07\textwidth
\begin{minipage}{0.63\linewidth}
\begin{center}
\epsfig{figure=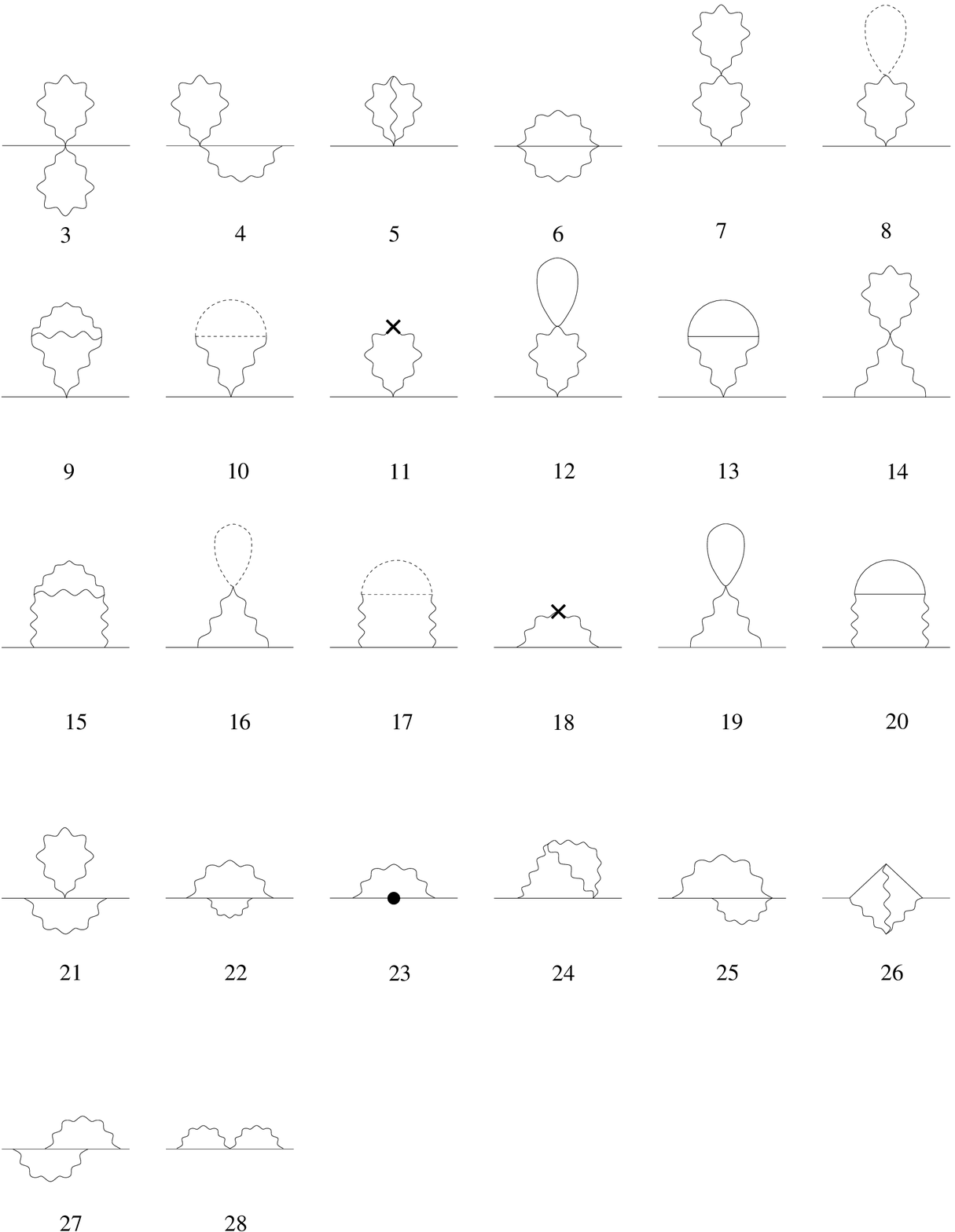,scale=0.5}
\end{center}
{\small {\bf Figure 2:} Two-loop diagrams
contributing to $dm_{\rm (2-loop)}$. Wavy (solid, dotted) lines
represent gluons (fermions, ghosts). Crosses denote vertices
stemming from the measure part of the action; a solid circle is a
fermion mass counterterm.}
\end{minipage}

The contribution of the $l^{{\rm th}}$ 1-loop diagram to $dm$, can
be expressed by:
\begin{equation}
d_l=\frac{(N^2-1)}{N}\,g^2 \cdot \sum_{i=0}^{2}c_{{\rm
    SW}}^i\,\varepsilon^{(i)}_{l}
\label{1loopContribution}
\end{equation}
where $\varepsilon^{(i)}_{l}$ are numerical constants and their
values depend on the Symanzik coefficients $c_i$. The dependence on
$c_{{\rm SW}}$ is seen to be polynomial of degree 2 ($i=0,\,1,\,2$).

The contribution of 2-loop diagrams without closed
fermion loops takes the form
\begin{equation}
d_l=\frac{(N^2-1)}{N^2}\,g^4 \cdot \sum_{i,j,k} c_{{\rm
SW}}^i\,N^j\,c_2^k\,e^{(i,j,k)}_{l}
\label{2loopContribution1}
\end{equation}
where the index $l$ runs over all contributing diagrams, $j=0,2$ and
$k=0,\,1,\,2$. The coefficients
$e^{(i,j,k)}_{l}$ exhibit a further dependence on Symanzik
coefficients (only through the combinations $C_1,\,C_2$), which
cannot be expressed in closed form and is presented
numerically in what follows.

The contribution of 2-loop diagrams,
containing closed fermion loops, has the form
\begin{equation}
d_l=\frac{(N^2-1)}{N}\,N_f\,g^4 \cdot
\sum_{i=0}^{4}c^{i}_{SW}\,\tilde{e}^{(i)}_{l}
\label{2loopContribution2}
\end{equation}

In order to enable cross-checks, numerical {\it
per-diagram} values of the constants $\varepsilon^{(i)}_{l}$,
$e^{(i,j,k)}_{l}$ and $\tilde{e}^{(i)}_{l}$ are presented in our
forthcoming publication~\cite{SCP}, for the Iwasaki action.

The total contribution of 1-loop diagrams, for $N=3$, can be written
as a polynomial function of the clover parameter $c_{{\rm SW}}$. For
the Wilson and Iwasaki actions we find, respectively:
\begin{equation}
dm_{\rm (1-loop)}^{\rm Wilson}
=g^2\,\Big(-0.434285489(1)+0.1159547570(3)\,c_{{\rm
SW}}+0.0482553833(1)\,c_{{\rm SW}}^2\Big)
\label{dm1loopWilson}
\end{equation}
\begin{equation}
dm_{\rm
(1-loop)}^{\rm Iwasaki}=g^2\,\Big(-0.2201449497(1)+0.0761203698(3)\,c_{{\rm
SW}}+0.0262264231(1)\,c_{{\rm SW}}^2\Big)
\label{dm1loopIwasaki}
\end{equation}

{\samepage{To illustrate our 2-loop results for some particular choices of the
action, we set $N=3$, $c_2=0$ and we use three different values for
the flavour number: $N_f=0,\,2$. Thus, for the Wilson action:
\begin{eqnarray}
N_f=0:&& \quad
dm_{\rm (2-loop)}=g^4\,\Big(-0.1255626(2)+0.0203001(2)\,c_{{\rm SW}}+0.00108420(7)\,c_{{\rm SW}}^2 \nonumber \\
&& \hspace{3.5cm} -0.00116538(2)\,c_{{\rm SW}}^3-0.0000996725(1)\,c_{{\rm SW}}^4\Big) \\
N_f=2:&& \quad
dm_{\rm (2-loop)}=g^4\,\Big(-0.1192361(2)+0.0173870(2)\,c_{{\rm SW}}+0.00836498(8)\,c_{{\rm SW}}^2 \nonumber \\
&& \hspace{3.5cm} -0.00485727(3)\,c_{{\rm
SW}}^3-0.0011561947(4)\,c_{{\rm SW}}^4\Big)
\label{dm2loopWilson}
\end{eqnarray}
and for the Iwasaki action:
\begin{eqnarray}
N_f=0:&& \quad
dm_{\rm (2-loop)}=g^4\,\Big(-0.0099523(2)-0.0024304(5)\,c_{{\rm SW}}-0.00232855(4)\,c_{{\rm SW}}^2 \nonumber \\
&& \hspace{3.5cm} -0.00032100(2)\,c_{{\rm SW}}^3-0.0000419365(1)\,c_{{\rm SW}}^4\Big)\\
N_f=2:&& \quad
dm_{\rm (2-loop)}=g^4\,\Big(-0.0076299(2)-0.0040731(5)\,c_{{\rm SW}}+0.00102758(6)\,c_{{\rm SW}}^2 \nonumber \\
&& \hspace{3.5cm} -0.00242924(3)\,c_{{\rm
SW}}^3-0.000457690(2)\,c_{{\rm SW}}^4\Big)
\label{dm2loopIwasaki}
\end{eqnarray}
\nopagebreak
In Figures 3 and 4 we present the values of $dm_{\rm (2-loop)}$ for
$N_f=0,\,2$, respectively; the results are shown for all our choices of
Symanzik actions, as a function of $c_{\rm
SW}\,(N=3,\,c_2=0)$.}}

\phantom{a}

\vskip -15mm
\noindent
\begin{minipage}{0.45\linewidth}
\begin{center}
\psfig{file=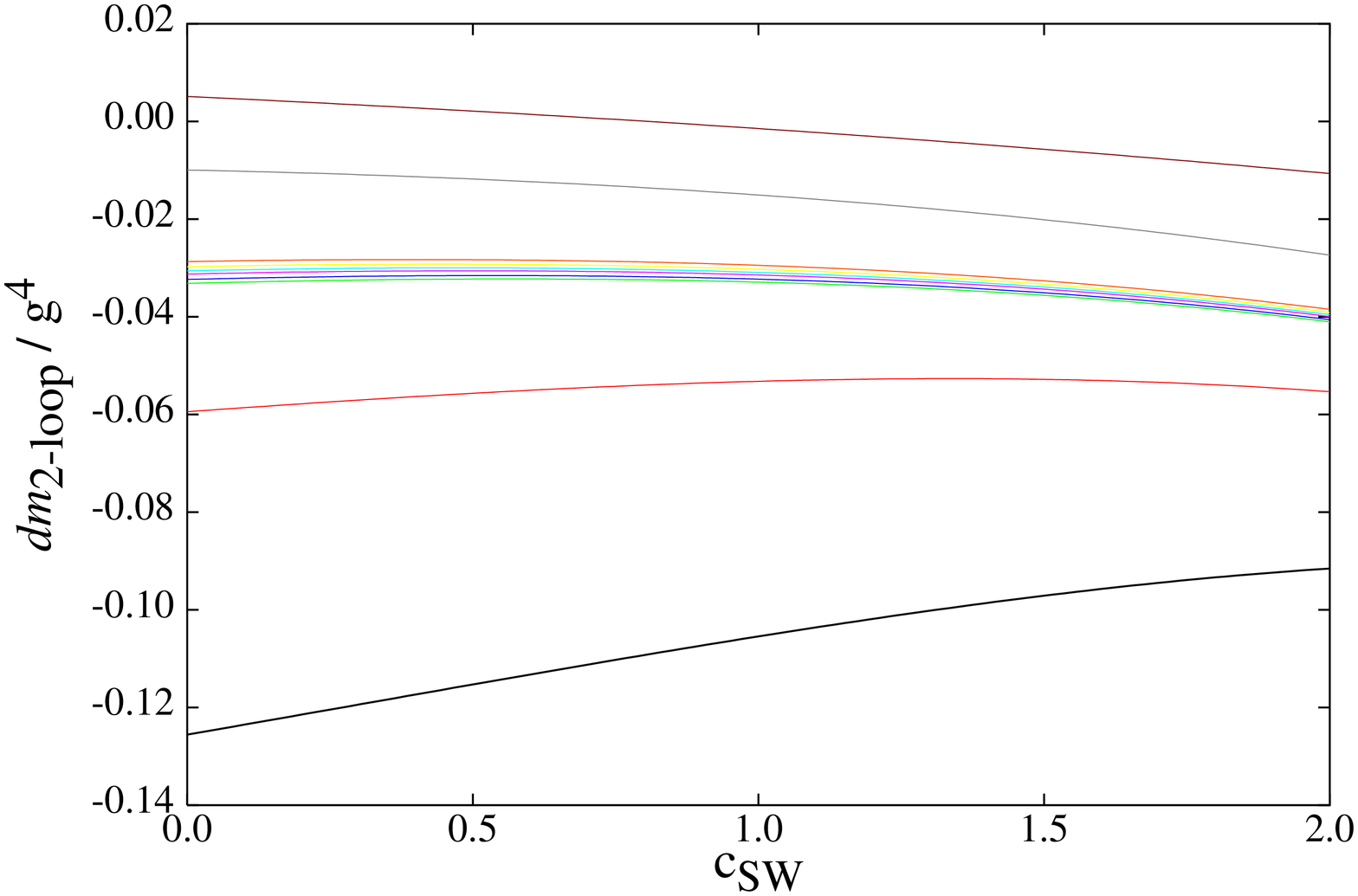,scale=0.27,angle=-90}
\end{center}
{\small {\bf Figure 3:} Total
contribution of 2-loop diagrams, for $N=3$, $N_f=0$ and $c_2=0$. Actions (top to bottom): \RawSienna{DBW2}, \Gray{Iwasaki}, TILW($\beta$=\RedOrange{8.00}, \Yellow{8.10}, \ProcessBlue{8.20}, \Magenta{8.30}, \Green{8.45}, \Blue{8.60}), \Red{Symanzik}, \Black{Plaquette}.}
\end{minipage}\hskip0.1\textwidth
\begin{minipage}{0.45\linewidth}
\begin{center}
\psfig{file=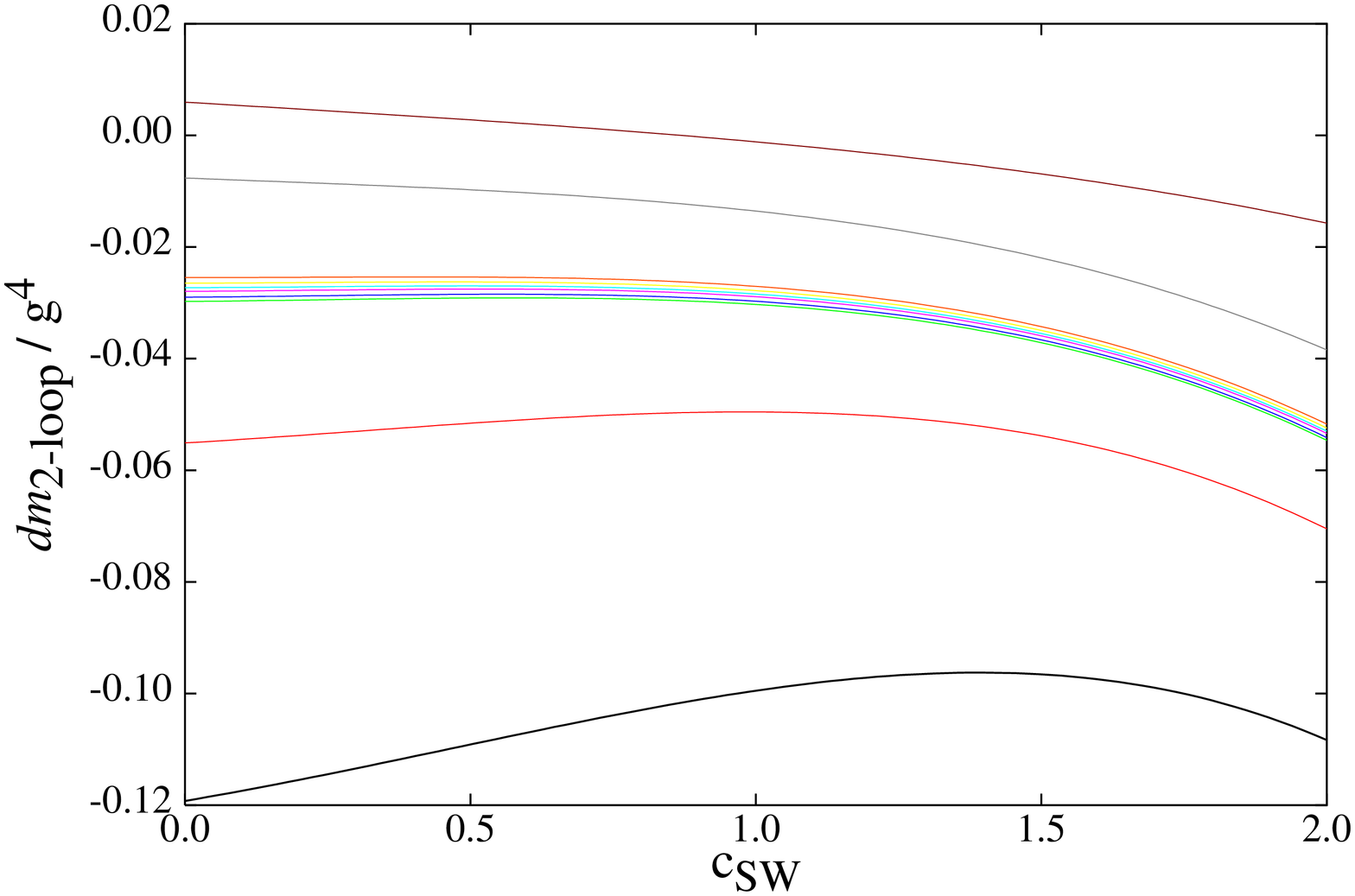,scale=0.27,angle=-90}
\end{center}
{\small {\bf Figure 4:} Total
contribution of 2-loop diagrams, for $N=3$, $N_f=2$ and $c_2=0$. Actions (top to bottom): \RawSienna{DBW2}, \Gray{Iwasaki}, TILW($\beta$=\RedOrange{8.00}, \Yellow{8.10}, \ProcessBlue{8.20}, \Magenta{8.30}, \Green{8.45}, \Blue{8.60}), \Red{Symanzik}, \Black{Plaquette}.}
\end{minipage}

\section{Improved Perturbation Theory}
\label{sec4}

We now apply our method of improving perturbation theory \cite{CPS},
based on {\bf resummation of an infinite subset of tadpole
diagrams}, termed ``cactus'' diagrams. In Ref. \cite{CPS} we show
how this procedure provides a simple way of dressing (to all orders)
perturbative results at any given order (such as the 1- and 2-loop
results of the present calculation). Some alternative ways of
improving perturbation theory have been proposed in Refs.
\cite{Parisi-81,L-M-93}. In a nutshell, our procedure involves
replacing the original values of the Symanzik and clover
coefficients by improved values, which are explicitly computed in
\cite{CPS}. Taking also due care to avoid double counting of
diagrams, we calculate the improved (``dressed'') value $dm^{\rm
dr}$ of the critical mass $(N=3,\,c_2=0)$.

{\samepage{We choose to study the case of the Wilson action ($\beta=5.29$) and
the Iwasaki action ($\beta=1.95$) with $N=3$ and $N_f=2$. Using
these values, the contribution to $dm^{\rm dr}_{\rm (2-loop)}$ is a
polynomial of $c_{\,{\rm SW}}$:
\begin{eqnarray}
dm^{\rm dr}_{\rm (2-loop),\,Wilson}=&-&0.77398(5)
+0.1632999(3)\,c_{{\rm SW}}+0.06225(2)\,c_{{\rm SW}}^2 \nonumber \\
&-&0.00440064(8)\,c_{{\rm SW}}^3 -0.000737797(1)\,c_{{\rm SW}}^4
\label{dm2loopDressedWilson}\\
dm^{\rm dr}_{\rm (2-loop),\,Iwasaki}=&-&0.0813302(9)+0.043030(3)\,c_{{\rm SW}}
+0.0308196(2)\,c_{{\rm SW}}^2 \nonumber \\
&-&0.00767090(8)\,c_{{\rm SW}}^3-0.001160923(1)\,c_{{\rm SW}}^4
\label{dm2loopDressedIwasaki}
\end{eqnarray}

\begin{table}[b]
\begin{minipage}{16cm}
\begin{center}
\begin{tabular}{lccr@{}lr@{}lr@{}lr@{}lr@{}lr@{}l}
\hline \hline
\multicolumn{1}{c}{Action}&
\multicolumn{1}{c}{$N_f$}&
\multicolumn{1}{c}{$\phantom{a}\beta\phantom{\Bigl(}$}&
\multicolumn{2}{c}{$c_{\rm SW}$}&
\multicolumn{2}{c}{$\kappa_{\rm 1-loop}$} &
\multicolumn{2}{c}{$\kappa_{\rm 2-loop}$} &
\multicolumn{2}{c}{$\kappa_{\rm 1-loop}^{\rm dr}$} &
\multicolumn{2}{c}{$\kappa_{\rm 2-loop}^{\rm dr}$} &
\multicolumn{2}{c}{$\kappa_{\rm cr}^{\rm non-pert}$} \\
\hline \hline
Plaquette &0  &6.00&  1&.479        &0&.1301      &0&.1335    &0&.1362    &0&.1362    &0&.1392\\
Plaquette &0  &6.00&  1&.769        &0&.1275      &0&.1306    &0&.1337    &0&.1332    &0&.1353\\
Plaquette &2  &5.29&  1&.9192       &0&.1262      &0&.1307    &0&.1353    &0&.1341    &0&.1373\\
Iwasaki   &2  &1.95&  1&.53         &0&.1292      &0&.1368    &0&.1388    &0&.1379    &0&.1421\\
\hline \hline
\end{tabular}
\caption{1- and 2-loop results, and non-perturbative estimates for
  $\kappa_{\rm cr}$
\label{tab14}}
\end{center}
\end{minipage}
\end{table}
\nopagebreak
The comparison between the total dressed contribution $dm^{\rm
dr}=dm^{\rm dr}_{\rm (1-loop)}+dm^{\rm dr}_{\rm (2-loop)}$ and the
unimproved contribution, $dm$, for the plaquette action
$(\beta=5.29,\,N_f=2)$ is exhibited in Figure 5, as a function of
$c_{\rm SW}$. Similarly, $dm$ for the Iwasaki action
($\beta=1.95,\,N_f=2$) is shown in Figure 6.
\nopagebreak
Finally, in Table 2, we present a comparison of dressed and
undressed results with non perturbative estimates for $\kappa_{\rm
cr}$ \cite{LSSWW,JS,UKQCD}. We observe that improved
perturbation theory, applied to 1-loop results, already leads to a
much better agreement with the non perturbative estimates.}}

\vskip 3mm
\noindent
\begin{minipage}{0.45\linewidth}
\begin{center}
\epsfig{file=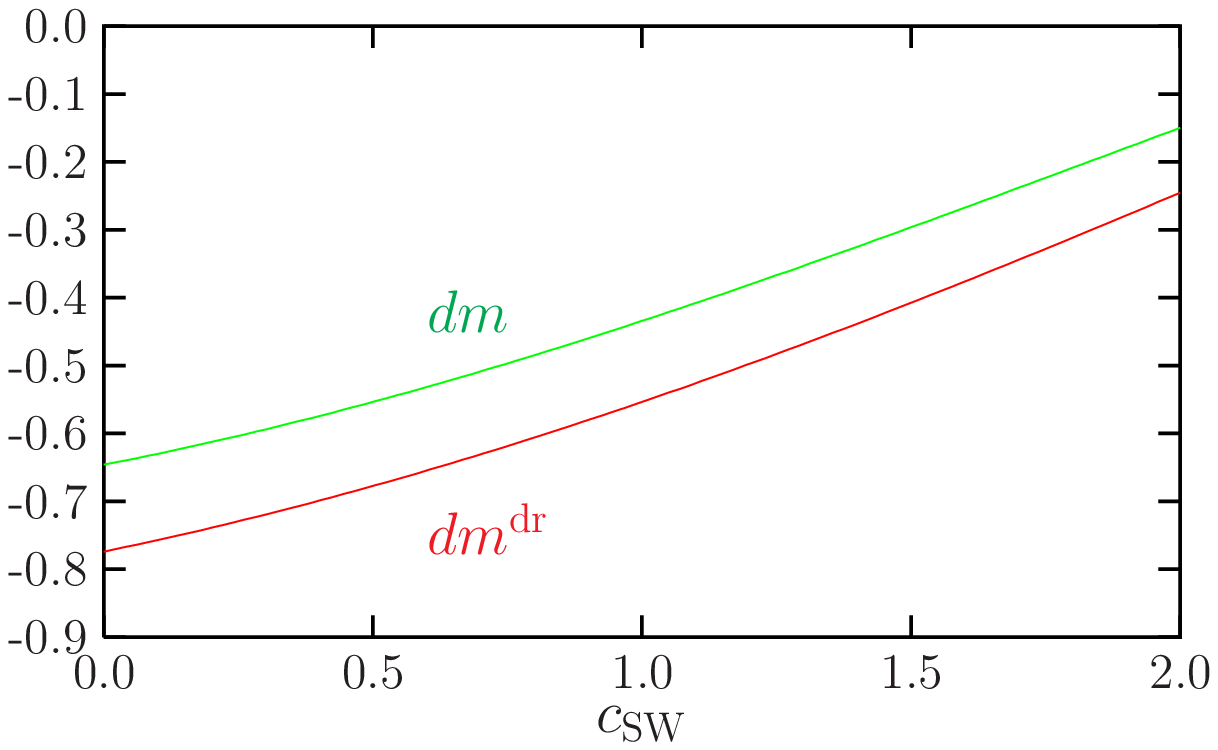,scale=0.5}
\end{center}
\vskip -2mm
{\small {\bf Figure 5:} Improved and unimproved values of $dm$ as a function of
$c_{\rm SW}$, for the plaquette action.}
\end{minipage}\hskip0.1\textwidth
\begin{minipage}{0.45\linewidth}
\begin{center}
\epsfig{file=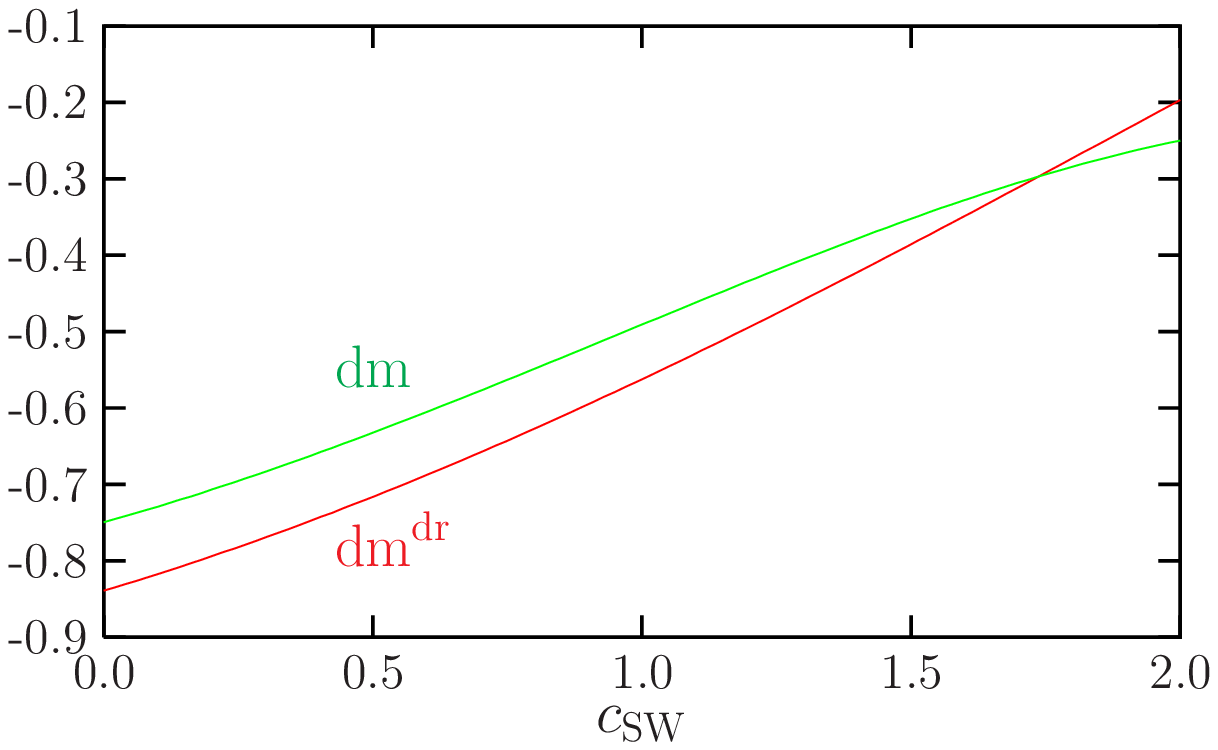,scale=0.5}
\end{center}
\vskip -2mm
{\small {\bf Figure 6:} Improved and unimproved values of $dm$
as a function of $c_{\rm SW}$, for the Iwasaki action.}
\end{minipage}
\vskip 3mm
\noindent
{\large \bf Acknowledgements: } Work supported in part by the 
Research Promotion Foundation of Cyprus.

\end{document}